\documentclass{iau-JDSS}
\usepackage{graphicx}

\title[I-GALFA] %% give here short title %%
{I-GALFA: The Inner-Galaxy ALFA Low-Latitude H I Survey}

\author[Koo et al.]   %% give here short author list %%
{Bon-Chul Koo$^1$, 
Steven J. Gibson$^{2,3}$, 
Ji-hyun Kang$^{1,2}$, 
Kevin A. Douglas$^{4,5}$,
Geumsook Park$^1$, 
Joshua E. G. Peek$^5$, 
Eric J. Korpela$^5$, 
Carl E. Heiles$^5$, 
Thomas M. Bania$^6$}
\affiliation{$^1$Seoul National Univ., KOREA;
$^2$Arecibo Obs., USA;
$^3$Western Kentucky Univ., USA; \\
$^4$Univ.\ of Exeter, UK;
$^5$Univ.\ of California - Berkeley, USA;
$^6$Boston Univ., USA }
\pubyear{2009}
\volume{Volume 15}  %% insert here IAU Highlights of Astronomy Volume Number
\pagerange{119}
\date{?? and in revised form ??}
\jname{Highlights of Astronomy, Volume 14}
\editors{Ian F Corbett, ed.}
\begin{document}

\maketitle

%%\begin{abstract}
%%\keywords{Keyword1, keyword2, keyword3, etc.}
%%\end{abstract}

%%\firstsection % if your document starts with a section,
%%\section{Introduction}

%%been examined (see for example \cite[Hwang \& Tuck 1970]{Hwang70};
%%\cite[Lee 1971]{Lee71}; \cite{Figer02}).

The I-GALFA survey is mapping all the {\sc H~i} in the inner Galactic disk
visible to the Arecibo 305m telescope within 10 degrees of the Galactic plane
(longitudes of $\ell=32^\circ$ to $77^\circ$ at $b=0^\circ$).  The survey,
which will obtain $\sim 1.3\times 10^6$ independent spectra, became possible
with the installation of the 7-beam Arecibo L-Band Feed Array (ALFA) receiver
in 2004.  ALFA's 
$3.\!'4$
%$3.'3 \times 3.'8$ 
resolution and tremendous sensitivity offer a great opportunity to observe the
fine details of {\sc H~i} in the Galaxy.  The I-GALFA survey began in 2008 May
and will be completed in 2009 September.  Night observations between May and
October are used for best spectral fidelity, allowing an RMS noise of $\sim
0.25$~K in 0.184 km~s$^{-1}$ channels covering LSR velocities of $-750$ to
$+750$ km~s$^{-1}$.
%
%%The total telescope time required is 310 hours spread over 130 nights. 
%The parameters of the survey are summarized in Table 1.
%%The mapping is done by ``basketweave'' scanning where 
%%the feed is driven up and down in zenith angle 
%%while the telescope is kept at the meridian. 
%%The spectra are obtained at every seconds by $\sim 8,000$-channel 
%%spectrometer, and these time-ordered data are 
%%later corrected for systematic instrumental effects. 
Details of the observing and data reduction can be found in \cite{Peek08}.  The
data will be made publicly available when the calibrated and gridded cubes are
completed.
%%It will be distributed through one or
%%more Virtual Observatory portals, including the Cornell Theory Center.
%%The data will be an essential material for the study of the  
%%diffuse interstellar meidum together with other major surveys.
Further information on the I-GALFA project 
may be found at {\bf www.naic.edu/$\sim$igalfa}.

%
%\begin{table}[h]\def~{\hphantom{0}}
%  \begin{center}
%  \caption{Parameters of the I-GALFA survey}
%  \label{tab:kd}
%  \begin{tabular}{ll}\hline
%      Mapping area & $|b|\le 10^\circ$, $\ell=32^\circ$ to 77$^\circ$ at $b=0^\circ$ (1157 square deg)\\
%      Telescope beam & $3.'3\times 3.'8$ \\
%      Spectral sampling & 0.184 km s$^{-1}$ \\ 
%      Spectral coverage & $-750$ to $+750$ km s$^{-1}$ \\ 
%      RMS noise & $\sim 0.25$ K \\\hline
%  \end{tabular}b
% \end{center}
%\end{table}
%

\vspace*{0.1in}

%\begin{figure}
\begin{center}
% iau09_fig1 image aspect: 1.8851 x 1.0000
%\includegraphics[height=3.7701in,width=2.0000in,angle=-90]{iau09_fig1.eps}
\begin{minipage}{5.3in}
\includegraphics[height=2.0in,width=2.0in,angle=0]{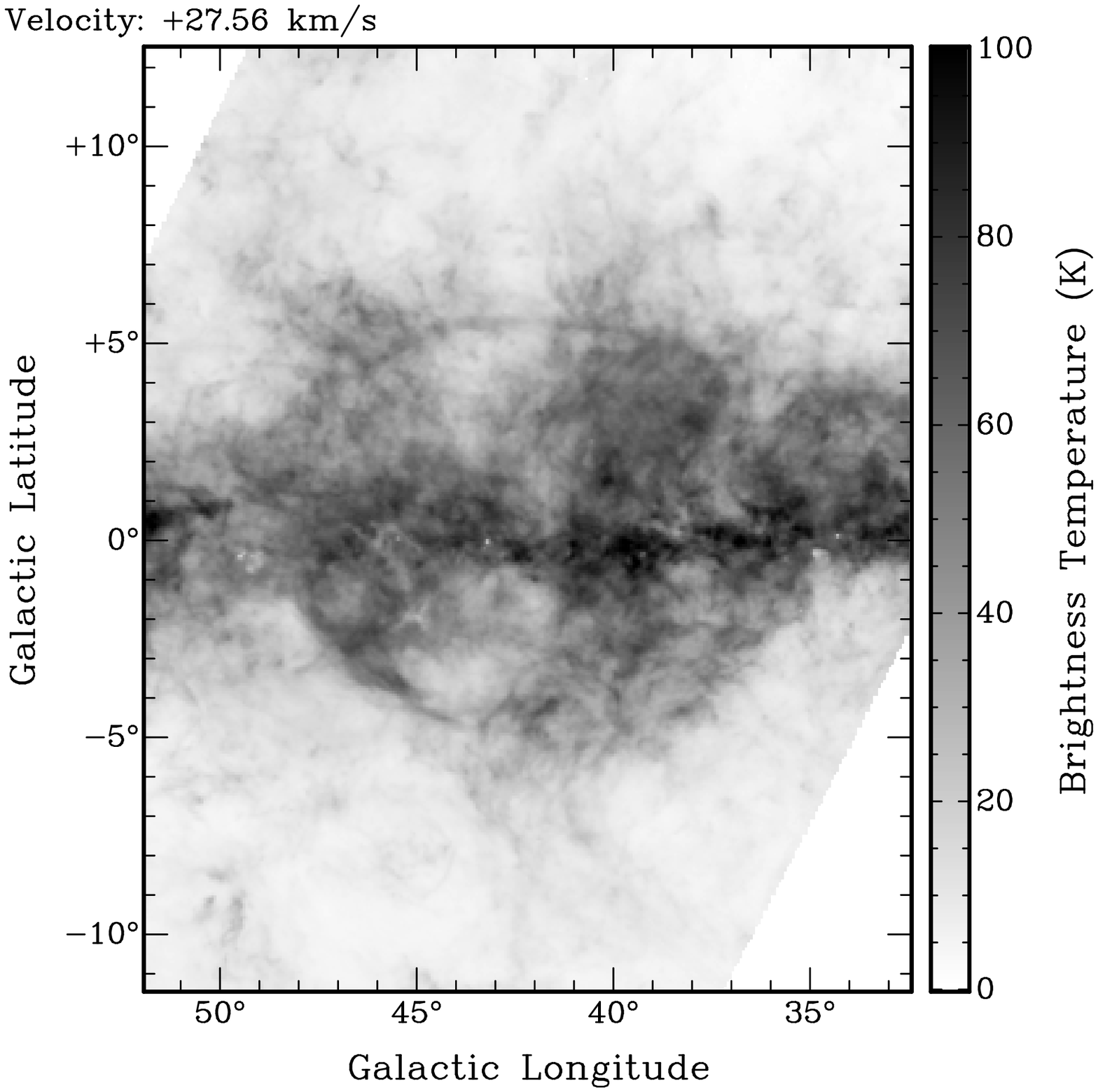}
~
\begin{minipage}{1.1in}
\vspace*{-2.0in}
{\bf Figure 1:}
Partial I-GALFA {\sc H~i} line channel maps; more data are being added. \\
{\it Left:} Supershell in the Sagittarius spiral arm. \\
{\it Right:} Disk-halo clouds, chimneys, and worms.
\end{minipage}
~
\includegraphics[height=2.0in,width=2.0in,angle=0]{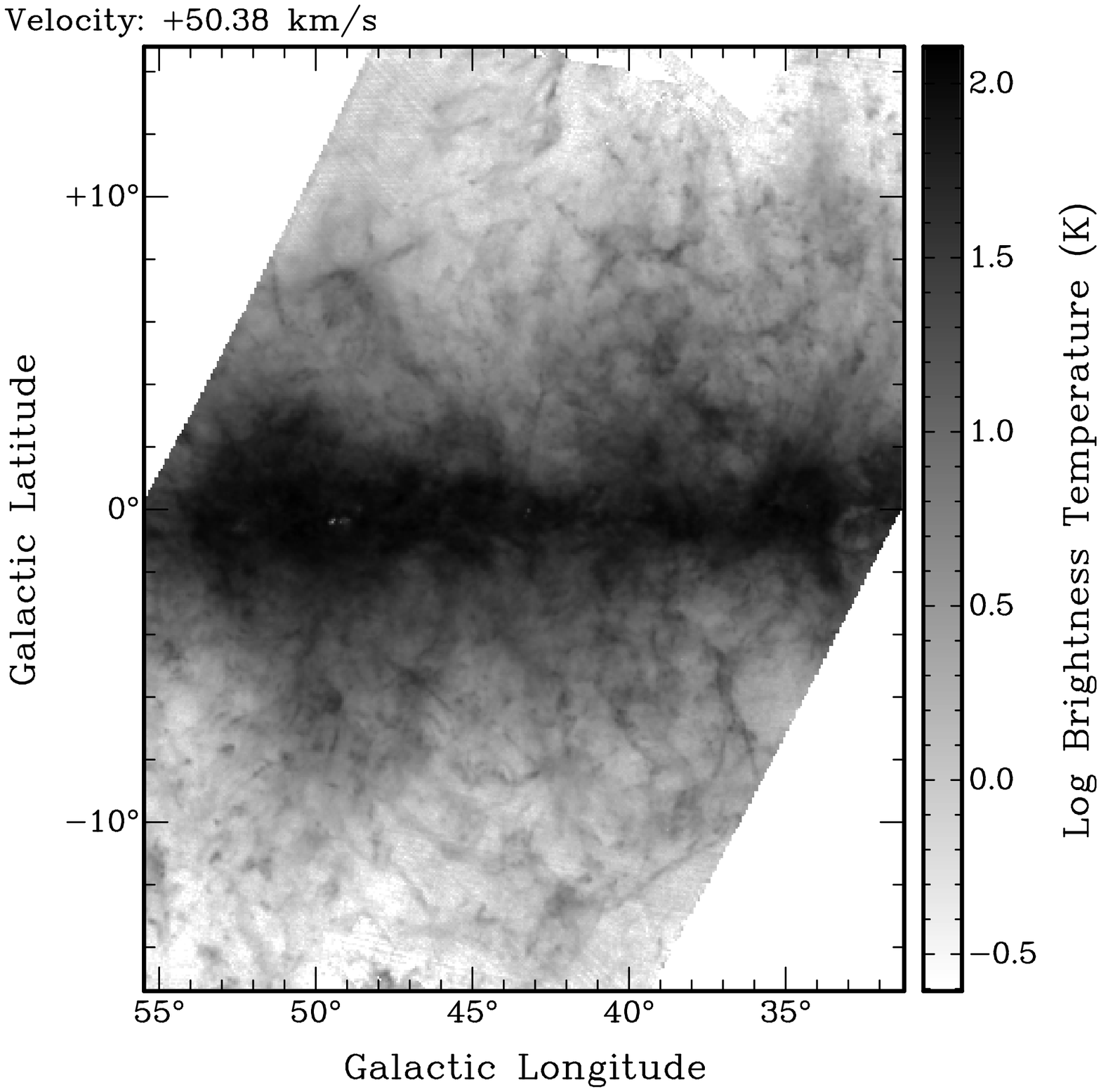}
\end{minipage}
\end{center}
%\caption{
%
%
%}\label{fig:hisa_weak_strong}
%\end{figure}

\begin{acknowledgments}
It is our great pleasure to thank all members of the AO staff for the support
of the I-GALFA survey.  B.-C.\ K. is supported by the Korean Research
Foundation under grant KRF-2008-313-C00372.  K. D. was supported by a Marie Curie
fellowship.  
%J.~P. was supported by NSF grants ??? and ???.  
The Arecibo
Observatory is part of the National Astronomy and Ionosphere Center, which is
operated by Cornell University under a cooperative agreement with the
U.S. National Science Foundation.

\end{acknowledgments}

%%\begin{discussion}

%%\end{discussion}


\begin{thebibliography}{}
%%
\bibitem[Peek \& Heiles (2008)]{Peek08}
     {Peek, J. E. G. \& Heiles, C. } 2008,
     \textit{astro-ph} arXiv:0810.1283v1

\end{thebibliography}
\end{document}